\documentclass[graybox]{svmult}

\usepackage{mathptmx}       % selects Times Roman as basic font
\usepackage{helvet}         % selects Helvetica as sans-serif font
\usepackage{courier}        % selects Courier as typewriter font
\usepackage{type1cm}        % activate if the above 3 fonts are
\usepackage{makeidx}         % allows index generation
\usepackage{graphicx}        % standard LaTeX graphics tool
\usepackage{multicol}        % used for the two-column index
\usepackage[bottom]{footmisc}% places footnotes at page bottom
\usepackage{epsfig}
\usepackage{amssymb}
\makeindex             % used for the subject index
\begin{document}

\title*{Study of Performance of Bakelite Resistive Plate Chamber (RPC)}
\author{R. Ganai, K. Agarwal, A.Roy, B. Muduli, S. Chattopadhyay, Z. Ahammed, G. Das, S. Ramnarayan}
\institute{R.Ganai, \at Variable Energy Cyclotron Centre, 1/AF, Bidhan Nagar, Kolkata-700064, \email{rajeshganai@tifr.res.in}
and \at K. Agarwal, \at Birla Institute of Technology and Science, Pilani, Rajasthan-333031, \email{f2011775@pilani.bits-pilani.ac.in}}
\maketitle

\abstract{Resistive Plate Chamber (RPC) is a type of gaseous detector having excellent time and position resolutions.
VECC is involved in the R\&D of indigenously developed bakelite RPCs. The largest size of bakelite RPC developed in India is 
100cm X 100cm.
We present here the test results of a bakelite sample along with the cosmic ray test results of a bakelite RPC (30cm X 30cm X 0.2cm) 
fabricated at VECC. The steps taken towards the development of a large size (240cm X 120cm X 0.2cm) bakelite RPC have also been 
discussed.
}

\section{Introduction}
\label{sec:1}
Resistive Plate Chamber (RPC) is a gas filled detector which utilises a constant and uniform electric
field produced between two high resistive parallel electrode plates e.g. glass, bakelite. 
Several high energy experiments like ALICE, CMS, ATLAS, BELLE use RPCs. RPCs can be used both for 
timing and trigerring purposes. RPCs will be used as active detectors in the proposed Iron CALorimeter (ICAL) experiment 
in India based Neutrino Observatory (INO) project \cite{poster},\cite{dae} and in the Near Detector (ND) of Deep Underground 
Neutrino Experiemnt(DUNE). ICAL will be a mammoth 50kTonnes of magnetised iron plates of dimension 48m X 16m X 0.05m stacked one
over another. RPCs of dimension $\sim$(2m X 2m X 0.002m) will be inserted in between two iron plates. There will be $\sim$150 such 
layers of iron plates.

\section{Fabrication of 30cm X 30cm X 0.2cm bakelite RPC}
\label{sec:2}
% Always give a unique label
% and use \ref{<label>} for cross-references
% and \cite{<label>} for bibliographic references
% use \sectionmark{}
% to alter or adjust the section heading in the running head
We have fabricated  \cite{thesis}  a single gap 30cm X 30cm X 0.2cm bakelite RPC from 0.3cm thick bakelite electrode.
The bakelite sheets were procured from local market.
A gap of 0.2 cm was maintained between these electrodes with the help of 4 button spacers, each of diameter 1 cm and height 
0.2 cm and 4 side spacers of length 28 cm and height 0.2 cm along with two gas nozzles. The outer surfaces of both the 
electrodes were painted with conducting black graphite paint mixed with special thinner in 1:1 ratio in order to apply 
uniform electric field across the plates. Figure. 1 shows the different components of the fabricated RPC.

\begin{figure}[]
\centering
%\sidecaption
% For example, with the graphicx style use
\includegraphics[width=10cm,height=10cm,keepaspectratio]{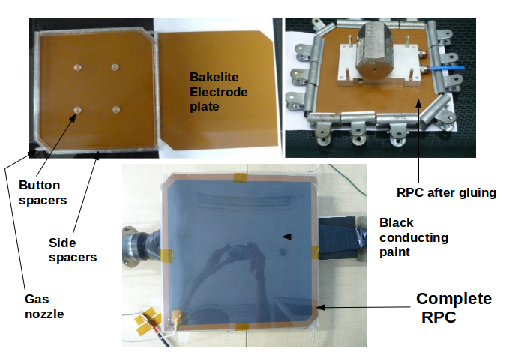}
%\texttt{sidecapion}
\caption{Components and different stages of fabrication of small size (30 cm X 30 cm X 0.2 cm) RPC}
\label{fig:1}       % Give a unique label
\end{figure}

\section{Results}
%The fabricated RPC has been tested in streamer mode. The gas composition in streamer mode is 
%Argon:Freon(R134a):Iso-butane::55:40:5. Before testing the RPC, the electrical properties of the bakelite sample has been tested.
\subsection{Electrical properties of bakelite sample}
\label{subsec:2}
The electrical properties like bulk resistivity and surface resistivity of the bakelite sample were measured and shown 
in figure 2. The bulk resistivity of the bakelite sample was measured to be $\sim$9X$10^{11}$$\Omega$cm whereas 
the surface resistivity was measured to be $\sim$3X$10^{12}$$\Omega$/$\square$.
\begin{figure}
\centering
%\sidecaption
\includegraphics[width=10cm,height=10cm,keepaspectratio]{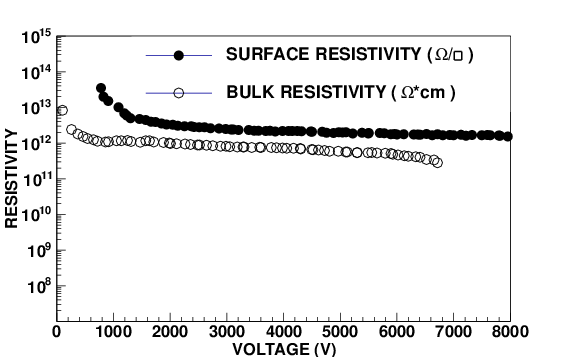}
\caption{Variation of bulk resistivity and surface resistivity with the applied total HV on the bakelite sample.}
\label{fig:2} 
\end{figure}

\subsection{I-V characteristics and current stability of RPC}
The I-V characteristics of the fabricated bakelite RPC has been studied in streamer mode of operation with a gas composition 
of Argon:Freon(R134a):Iso-butane::55:40:5 as shown in figure 3(a). Two distinct slopes in the I-V characteristics have been 
obtained with a breakdown voltage $\sim$7000V. The RPC has been tested for current stability for $\sim$50hrs at $\pm$6000V 
as shown in figure-3(b).
\begin{figure}
\centering
%\sidecaption
\includegraphics[width=13cm,height=13cm,keepaspectratio]{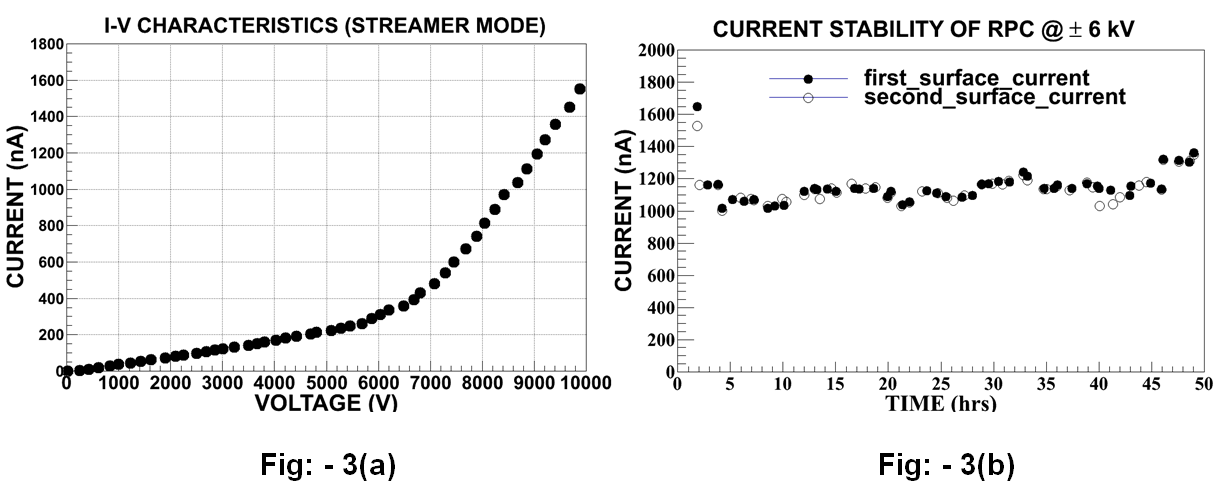}
\caption{Fig. 3(a) shows the I-V characteristics of the bakelite RPC and Fig. 3(b) shows the current stability of the bakelite
RPC at $\pm$6kV.}
\label{fig:3} 
\end{figure}

\subsection{Efficiency and noise rate}
The efficiency and noise rate of the RPC with cosmic rays have been was measured. An efficieny plateau was obtained for the RPC beyond 8000V. 
The efficiency, as shown in figure-4(a) was found to be $\sim$98$\%$. During this test, noise rate of the RPC was also calculated 
and was found to be $\sim$1.7 Hz/$cm^{2}$ at 9000V. A linearly varying behaviour of the noise rate with the applied voltage is 
shown in figure.4(b).

\begin{figure}
\centering
%\sidecaption
\includegraphics[width=12cm,height=12cm,keepaspectratio]{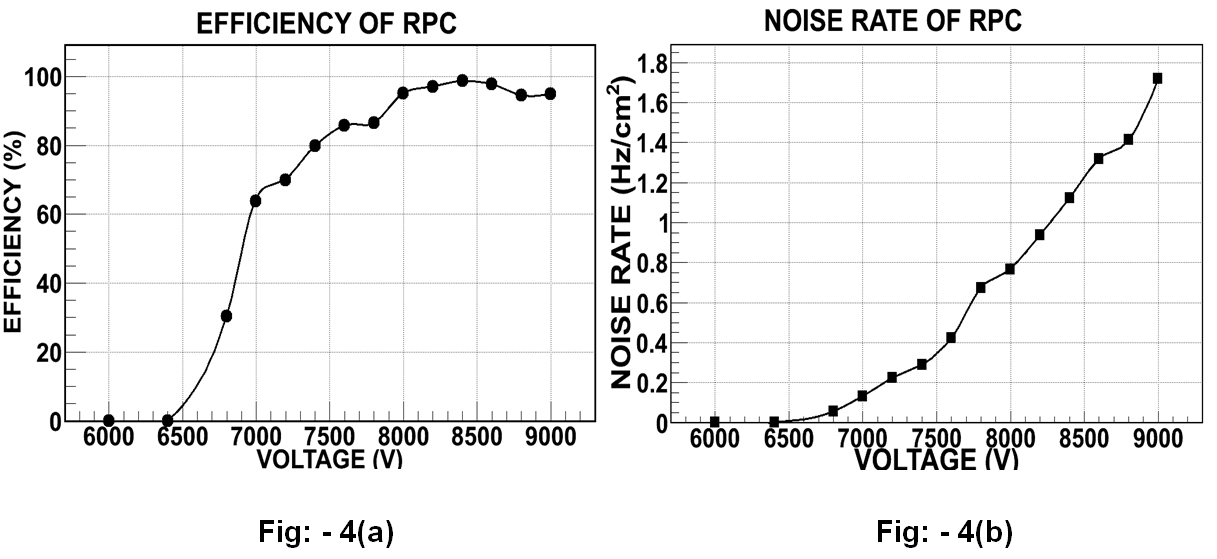}
\caption{Fig. 4(a) shows the efficiency and Fig. 4(b) shows the noise rate of fabricated the bakelite RPC and.}
\label{fig:4} 
\end{figure}

\section{Fabrication of large bakelite RPC}
With the successful results of the samll (30cm X 30cm X 0.2cm) bakelite RPC and acceptable values of bulk and surface resistivity 
of the bakelite sample, we have started to develope large size (240cm X 120cm X 0.2cm) bakelite RPC. Figure 5(a) shows the 
various steps towards the fabrication of such a large size RPC and figure 5(b) shows the large RPC under test.

\begin{figure}[tmb]
\centering
%\sidecaption
\includegraphics[width=10cm,height=10cm,keepaspectratio]{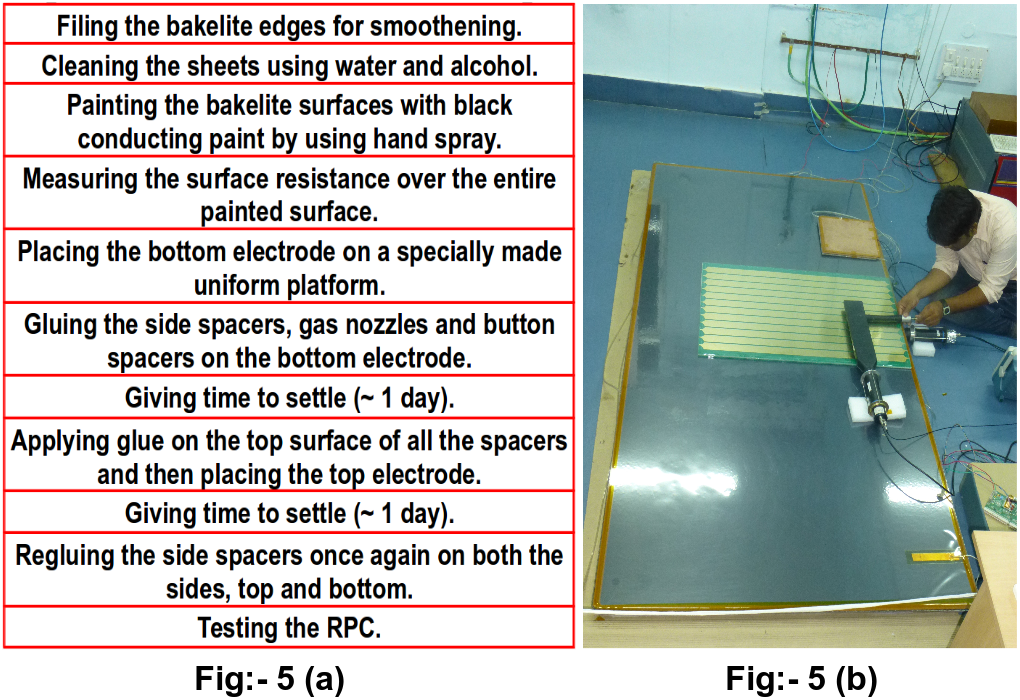}
\caption{Steps of fabrication of large bakelite RPC.}
\label{fig:5} 
\end{figure}

\section{Conclusion}
To conclude, the electrical properties of the indigenously produced bakelite sheets have been found promising for RPC fabrication. 
A single gap bakelite RPC of dimension 
30 cm X 30 cm X 0.2 cm has been fabricated and tested with cosmic rays for current stability, efficiency and noise rate. 
The results satisfy the use of this bakelite to fabricate RPCs for both ICAL and DUNE-ND. We are in the process of measuring 
the time resolution of the detectors. We have also developed large size (240cm X 120cm X 0.2cm) RPC from the same sample. The test of 
this RPC is under process.


\begin{thebibliography}{99.}%
% Use the following syntax and markup for your references if 
% the subject of your book is from the field 
% "Mathematics, Physics, Statistics, Computer Science"
%
% Contribution 

\bibitem{poster} R. Santonico R.Cardarelli, Nucl. Inst. and Meth. 187, (1987) 331.
\bibitem{dae} INO Project Report, INO/2006/0, May 2006.
\url{http://www.ino.tifr.res.in/ino}
\bibitem{thesis} Saikat Biswas, PhD Thesis, University of Calcutta, 
\url{http://www.ino.tifr.res.in/ino//theses.php}
\end{thebibliography}
\end{document}